\documentstyle[epsfig]{l-aa}

\input{psfig}

\begin{document}

\thesaurus{10.08.1, 10.11.1, 10.19.2, 12.04.1, 12.07.1}

\title{ Observational Limits on Machos in the Galactic Halo.
\thanks{Based on observations made at the European Southern Observatory,
La Silla, Chile.}}
\author {
C.~Renault\inst{1},~C.~Afonso\inst{1},~\'E.~Aubourg\inst{1},~P.~Bareyre\inst{1},~F.~Bauer\inst{1},~S.~Brehin\inst{1},~C.~Coutures
\inst{1},~C.~Gaucherel\inst{1},
 J.F.~Glicenstein\inst{1}, B.~Goldman\inst{1}, M.~Gros\inst{1}, D.~Hardin\inst{1}, J.~de~Kat\inst{1},
M.~Lachi\`{e}ze-Rey\inst{1},
B.~Laurent\inst{1}, \'E.~Lesquoy\inst{1},
C.~Magneville \inst{1}, A.~Milsztajn \inst{1},  L.~Moscoso\inst{1}, N.~Palanque-Delabrouille\inst{1}, 
F.~Queinnec\inst{1}, 
J.~Rich\inst{1}, M.~Spiro\inst{1},
L.~Vigroux\inst{1}, S.~ Zylberajch\inst{1},
R.~Ansari\inst{2}, F.~Cavalier\inst{2}, F.~Couchot\inst{2}, B.~Mansoux\inst{2}, M.~Moniez\inst{2}, O.~Perdereau\inst{2},
J.-Ph.~Beaulieu\inst{3},  R.~Ferlet\inst{3},  Ph.~Grison\inst{3}, A.~Vidal-Madjar\inst{3},
J.~Guibert \inst{4}, O.~Moreau\inst{4},  
\'E.~Maurice\inst{5}, L.~Pr\'{e}v\^{o}t\inst{5},
C.~Gry\inst{6},
S.~Char\inst{7}, J. ~Fernandez\inst{7}\\   \indent   \indent
The EROS collaboration
}
\institute{
CEA, DSM, DAPNIA,
Centre d'\'Etudes de Saclay, F-91191 Gif-sur-Yvette Cedex
\and
Laboratoire de l'Acc\'{e}l\'{e}rateur Lin\'{e}aire,
IN2P3 CNRS, Universit\'e Paris-Sud, F-91405 Orsay Cedex
\and
Institut d'Astrophysique de Paris, CNRS,
98~bis Boulevard Arago, F-75014 Paris
\and
Centre d'Analyse des Images de l'INSU,
Observatoire de Paris,
61 avenue de l'Observatoire, F-75014 Paris
\and
Observatoire de Marseille,
2 place Le Verrier, F-13248 Marseille Cedex 04
\and
Laboratoire d'Astronomie Spatiale de Marseille,
Traverse du Siphon, Les Trois Lucs, F-13120 Marseille
\and
Universidad de la Serena, Faculdad de Ciencias, Departemento de Fisica,
Casilla 554, La Serena, Chile
}

\offprints{cecile.renault@cea.fr}

\date{Received;accepted}
\maketitle
\markboth{Renault et al: Observational Limits on Machos in the Galactic Halo}{}

\begin{abstract}
\indent
We present final results from the first phase of the EROS search
for gravitational microlensing of stars in the 
Magellanic Clouds by unseen deflectors (machos: MAssive Compact Halo Objects).
The search is sensitive to events with time scales between 15 minutes and 200 days corresponding to deflector masses
in the range 10$^{-7}$ to a few M$_\odot$.
Two events were observed that are compatible with microlensing by objects of mass $\approx$ 0.1 M$_\odot$.
By comparing the results with the expected number of events for various models of the Galaxy, 
we conclude that machos in the mass range [$ 10^{-7},\ 0.02$]  M$_\odot$ make up less than 20~\% (95 \% C.L.)
of the Halo dark matter.

\end{abstract}

\section{Introduction}
\indent
The presence of large quantities of ``dark matter'' in
spiral galaxies like our own has
been inferred from their flat rotation curves (e.g. Primack {\it et al}, 1988); 
the dynamic mass of the Galaxy is thought to be 3 to 8 times larger than the visible mass up to 50 kpc from the Galactic Center.
From primordial nucleosynthesis, we learn that baryonic dark matter can be up to 10 times more abundant 
than visible matter (depending on the value of $H_0$): all Galactic dark matter could thus be baryonic.
A possible form would be compact objects too light to burn hydrogen
\mbox{(m $ < 0.07 - 0.1$ M$_\odot$)} (Carr, 1990).
\\   \indent
Here we report results from a search for such unseen
compact objects in the Galactic Halo
performed by our collaboration ``EROS''
(Exp\'{e}rience de Recherche d'Objets Sombres)
at the European Southern Observatory at La Silla,
Chile.
Such objects can be detected via
the gravitational microlensing effect (Paczy\'nski, 1986)
that causes an apparent temporary brightening of stars outside
our Galaxy as the unseen object passes near the line of sight.
The magnification factor  is given by 
\mbox{$A=(u^{2}+2)/[u(u^{2}+4)^{1/2}]$}
where $u$ is the undeflected ``impact parameter'' of the light
ray with respect to the unseen object in units of the ``Einstein
Radius'',  \mbox{$R_{E}=(4GmLx(1-x)/c^{2})^{1/2}$.}
Here, $m$ is the deflector
mass, $L$ the observer-source distance and $Lx$ the
observer-deflector distance.
\\   \indent
The time scale $\tau$ for
the magnification is the time for a Halo object to move through a
distance equal to its Einstein radius; its median value is
\mbox{$\tau \sim$ 90 days $\sqrt{m/M_\odot}$}
for $L$= 55~kpc (source star in the LMC) and a standard model of the Halo with a velocity dispersion of \mbox{245~km.s$^{-1}$.}
\\   \indent
EROS has conducted two observing programs, one using a
16 CCD camera mounted on a 40 cm diameter telescope
to search for short time scale microlensing events \mbox{($\tau <$ a} few days),
and the other using Schmidt photographic plates for longer time scales.
Results from the first 2 years of  CCD data were given in (Aubourg {\it et al}, 1995) while results
from all 3 years of Schmidt plate data were published in (Ansari {\it et al}, 1996).
In this paper, we present final results from the first phase of the EROS program, corresponding to 4 years of CCD observations
and the Schmidt plate data.

\section{Observations and data reduction}

\indent
During 3 annual periods of about 6 months, 290 usable photographic
plates of $29\times29$ cm$^2$ have been exposed at the ESO
1m Schmidt telescope, half with a red filter and half with a blue filter.
The usable field on a plate is $5.2^{\circ}\times5.2^{\circ}$, centered
on \mbox{$\alpha$ = 5h20mn}, \mbox{$\delta = -68^{\circ}30'$} (eq.~2000).
Exposure times were 1 hour in each colour. Apart from the
very crowded LMC bar region,
our star detection efficiency abruptly drops at limiting
magnitudes of about 20.5 in red and 21.5 in blue.
The time sampling of the plates makes the program sensitive to
microlensing event durations ranging from one day to a few months.
Details are available in (Cavalier, 1994), (Laurent, 1995), (Ansari {\it et al}, 1996).
\\   \indent
We have also taken during four annual periods more than 19,000 images with our CCD setup.
The camera  (Arnaud {\it et al}, 1994) consisted of a mosaic of 16 buttable \mbox{579 $\times$ 400} pixels \mbox{Thomson THX 31157} CCDs.
It was mounted on a 40 cm reflector (f/10) refurbished by us and the Observatoire de Haute-Provence 
and had a field of $0.4^{\circ} \times 1.1^{\circ}$. 
Eleven CCDs were active in \mbox{1991-92} during 100 days and 15 CCDs for the next three seasons of about 230 days each. 
The first three years were devoted to the observation of one field in the bar
of the LMC
\mbox{$(\alpha$ = 5h23.5mn}, \mbox{$\delta = -69^{\circ}36'$}, eq.~2000),
 the last year to one field in the center of the SMC
\mbox{$(\alpha$ = 0h50mn}, \mbox{$\delta = -73^{\circ}15'$}, eq.~2000). Here, we are sensitive to
microlensing durations ranging from 15 minutes to a few days on stars brighter than about 19.5 magnitude in V band.
Details are available in (Queinnec, 1994), (Aubourg at al, 1995), (Renault, 1996a), (Renault {\it et al}, 1996b).
\\   \indent
The images are processed using a custom designed fast photometric reconstruction software to produce light curves. We then analyse
the 250 000 light curves from the CCD data and the 6~10$^6$ light curves from Schmidt plates in both colours,
 corresponding to a total of 5~10$^9$ photometric measurements in crowded fields.

\section{Data analysis}

\indent
In order to isolate microlensing events, we searched for positive variations  occuring simultaneously in the red and blue light
curves. CCD and Schmidt plate analyses differed in details but used the same  properties which
allow us to separate microlensed stars from variable stars.
The most important one is
uniqueness: because of the low predicted optical depth \mbox{($\stackrel{<}{_\sim} 5.10^{-7}$)}, 
the probability of a measurable microlensing effect to happen twice on the same star
in a few years time  is negligible.
We also require equality of the base luminosity before and after the variation.
In addition to these properties, the set of microlensing candidates should be representatively distributed
 throughout our observed H-R diagram.
Neither analysis uses the shape of the light curve, thus making us sensitive to events involving multiple lenses or sources, and
to events affected by the finite source size or additional light from unresolved stars (blending).
\\   \indent
To obtain detection efficiencies, we processed through the same analysis software a sample of observed light curves
 with a random theoretical microlensing shape superimposed; blending and finite source size effects are
 taken into account.
The finite size effect results essentially in a large loss of efficiency for 
lensing objects lighter than $10^{-6}$ M$_\odot$; it is negligible for deflectors heavier than $10^{-4.5}$~M$_\odot$
(Renault {\it et al}, 1996b). 
The blending correction is about four times smaller than that due to the finite size of 
the source for the CCD data. Its impact on the Schmidt plates analysis 
is not important (Ansari {\it et al}, 1996).
\\   \indent
No microlensing event was identified in the CCD data.
Two events compatible with microlensing were
identified from the Schmidt plate data. In the hypothesis of microlensing, they have amplitudes of 1.0 and 1.1 
mag.
and time scales of 23 and 29~days. Those durations correspond to deflectors in the mass range
[0.01-1~M$_\odot$].
The light curves were presented in (Aubourg {\it et al}, 1993), (Ansari {\it et al}, 1996).
We have taken several spectra and followed the candidates with our CCD camera.
The first candidate is a Be star (Beaulieu {\it et al}, 1995) and has shown no subsequent variation in plate or CCD observations 
over a 3 year period.
The second candidate is an A0 star
that exhibits a 2.8 day periodic variability suggestive of an eclipsing system (Ansari {\it et al}, 1995).
However, the observed single large increases ($\approx$ 1 mag.) do not correspond to any known variability 
of such stars;
 we cannot exclude that the detected events are due to new types of variable star.

\section{Expected number of events}

\indent
The expected number of events can be calculated once a model of the Galaxy is chosen.
Several models were simulated using a combination of a disk and a halo. 
The disk models are exponential with a height scale of 0.5~kpc and a length scale of 5.0~kpc. We use a surface density $\Sigma_0$ of 
50 or 100~M$_\odot $pc$^{-2}$ ; the lowest value corresponds to known matter in the disk 
(stars, stellar remnants and gas), whereas the largest value is
suggested by maximal disk models and includes a dark matter component in the disk.
\\   \indent 
For the Halo, we have simulated spherical ``standard" halos and flattened halos (Evans, 1993).
Flattened halo models are self-consistent and give simultaneously density and velocity dispersion.
The flattening is characterized by the axis ratio $q$ and the shape of the rotation curve by a parameter $\beta$;
 we assume $q$ = 0.75 and an asymptotically flat rotation curve \mbox{($\beta=0$)}.
For spherical and flattened models, we use a core radius of 5.6~kpc (Primack {\it et al}, 1988)
and a Galactic Center distance of 7.9~kpc (Merrifield, 1992).
We scale the halo density distributions such that
 the rotation velocity near the Sun be 200~km.s$^{-1}$ (Merrifield, 1992). We have also simulated our
``reference model" used in (Ansari {\it et al}, 1996). It is
 a ``standard" spherical model with a core radius of 7.8~kpc and a velocity dispersion of 245~km.s$^{-1}$.
The mass of its Halo is normalised to 4.10$^{11}$~M$_\odot$ within 50~kpc. Table \ref{tab} gives 
characteristics for five models: the local Disk surface density, the local Halo density and
 the circular velocity at 50~kpc.
Other models are described in (Renault {\it et al}, 1996b).

\vspace{-0.35cm}
\begin{table}[hhh]
\begin{center}
\vfill
 \begin{tabular}{|c|c|c|c|c|c|}
\hline
 & shape & $\Sigma_0$                 &   $\rho_\odot$          & \multicolumn{2}{|c|}{$V_{circ}$ at 50 kpc}   \\
  & of the &   (M$_\odot $pc$^{-2})$  &   (M$_\odot $pc$^{-3}$) & \multicolumn{2}{|c|}{(km.s$^{-1}$)}  \\
 &  halo     &  & &  Halo & total                               \\
\hline
\hline
1 & spherical & 50  &  0.017 & 269  & 275  \\
\hline
2 & spherical &   100  &   0.010  & 203  & 220  \\
\hline
3 & flattened &  50  &    0.020 &  204 &  212  \\
\hline
4 & flattened &   100  &   0.012 & 154  & 175   \\
\hline
 & (reference) &       &   &     &   \\
5 & spherical &  -     &  0.008  & 185    & -  \\
\hline
\hline
\end{tabular}
\end{center}
\normalsize
\caption{Characteristics of the five models; model 5 is the reference model described in  (Ansari {\it et al}, 1996).
$\Sigma_0$ is the local Disk surface density and $\rho_\odot$ is the local Halo density.
The circular velocities $V_{circ}$ are those generated by the mass in the Halo or by the total mass (disk+halo) at a radius equal to
 the LMC distance.}
\label{tab}
\vfill
\vspace{-0.35cm}
\end{table}

\begin{figure}
 \begin{center}
  \mbox{\vspace{-0.35cm}
\epsfig{file=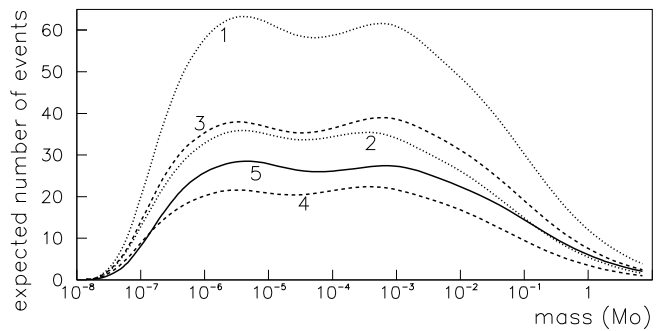,width=9.cm}}
     \caption{Expected number of microlensing events
              in the EROS programs (CCD -LMC and SMC sources- 
and Schmidt plates) assuming that all Galactic dark matter is in the form
of machos of the same mass. The five curves refer to the Galaxy models of table \ref{tab}.
The full curve corresponds to the reference model,
 dashed curves to flattened models and dotted curves to the spherical models.}\label{fig_nexp}
  \end{center}
\vspace{-0.35cm}
\end{figure}

  \indent
The expected number of events due to Halo objects as a function of their mass is plotted in figure~\ref{fig_nexp},
where it is assumed that all machos have the same mass $m$.
We have combined Schmidt plates and CCD programs.
Two comparable maxima are seen where observations are most efficient: at $m~=~3\ 10^{-6}$~M$_\odot$
 with the CCD data and  $m~=~10^{-3}$~M$_\odot$ with the Schmidt plates. 
The expected numbers of events for the five models approximately scale as $V_{circ}^2$ (table \ref{tab})
and therefore reflect the halo masses.

\indent
The expected number of events due to lensing by Galactic stars is less than 0.15; for lensing by 
stars in the Magellanic Clouds, it is less than 0.6.

\section{Constraints on the Halo}
\begin{figure}[hhh]
  \begin{center}
  \mbox{\vspace{-0.35cm}
\epsfig{file=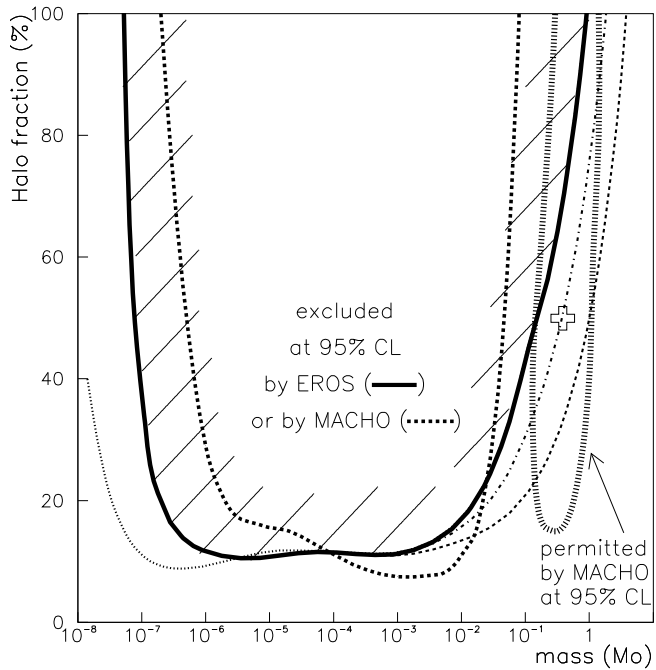,width=9.cm}}
    \caption{Exclusion diagram at 95~\%~CL for the reference model with all EROS data,
 assuming all deflectors to have the same mass. 
For the CCD program, we show the influence of blending and finite size effects 
(the dotted line on the left is the limit without those effects).
Limits are shown for 0 (dashed line), 1 (mixed line) or 2 (full line)
candidates assumed to be actual microlensing. The cross is centered on the area allowed at 95~\%~CL 
by the MACHO program (Alcock {\it et al}, 1996b) assuming 6 microlensing events and a standard spherical model (model S,
 very similar to our reference model). We also indicate the MACHO exclusion contour obtained by combining all their results
with $\tau \leq 20$ days (Alcock {\it et al}, 1996a). }\label{fig_exclupart}
  \end{center}
\vspace{-0.35cm}
\end{figure}

\indent
From the expected and observed number of events, we obtain 
an upper limit on the fraction of the halo along the line of sight that is composed of machos. The statistical method is the same 
as in (Ansari {\it et al}, 1996).
Figure \ref{fig_exclupart} presents 95~\%~CL limits for the reference model, all the EROS data
and two microlensing candidates.  This limit is shown as a function of the assumed deflector mass.
For the CCD data, we also show the effect of taking into account 
blending and finite size effects, important only below 10$^{-6}$~M$_\odot$; these effects
supress all sensitivity below 10$^{-7}$~M$_\odot$.
We also show limits assuming 0 or 1 
 microlensing events, that are identical for objects with  masses below
10$^{-3}$~M$_\odot$. For two microlensing events,
we can not give significant constraints beyond 1~M$_\odot$. 

\begin{figure}[hhh]
  \begin{center}
   \mbox{\vspace{-0.35cm}
\epsfig{file=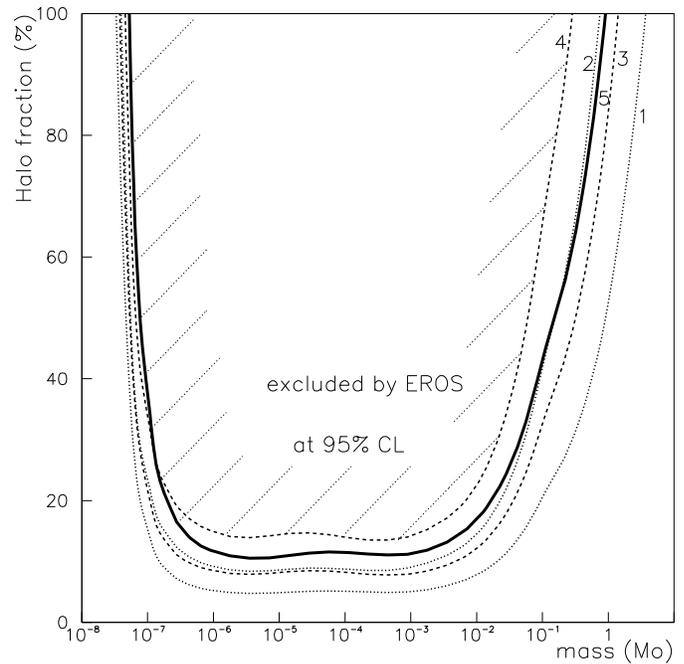,width=9.cm}}
    \caption{Exclusion diagram at 95~\%~CL assuming the same mass for all deflectors, 
for 5 Galaxy models of table \ref{tab}. The full curve corresponds to the reference model,
 dashed curves to flattened models and dotted curves to the spherical models.
 We have used all EROS data and assumed two observed microlensing events with $\tau$~=~23 and 29 days.}
  \label{fig_exclu}
  \end{center}
\vspace{-0.35cm}
\end{figure}

Figure \ref{fig_exclu} presents the limits obtained for different models of the Galaxy and all the EROS data.
 The hypothesis that the two observed events are true microlensing events is adopted in order to obtain a 
conservative upper limit.
\\
\indent
The used disk length scale (DLS) is likely to be high (Sackett, 1996) but models with low disk surface density
 and a DLS of 2.5~kpc are more constraining than our models with a heavy disk
and a DLS of 5~kpc. Moreover, heavy disk and low DLS ($<$3.2~kpc) are not compatible with the observed
rotation curve.
\\
\indent
 We can divide the studied mass range [10$^{-8}$-1~M$_\odot$] in three distinct parts.
\\
1) No limits can be inferred for objects lighter than 10$^{-7}$~M$_\odot$;
 the very existence of such objects is doubtful.
\\   
2) For an intermediate model, we observe that
the fraction of the Halo in the form of objects with masses between  10$^{-7}$ and 0.02~M$_\odot$ is below 20~\%
(between 5~10$^{-7}$ and 0.002~M$_\odot$, it is lower than 10~\%). 
As the limit is rather mass independent, it is valid for any mass function in this interval.
This limit is similar to that of the MACHO collaboration (Alcock {\it et al}, 1996a) but extends an order of magnitude lower in mass.
We note that the limits from the two groups are derived from almost independent sets of stars observed at different times.
 Thus they could be combined to obtain an even lower limit.
\\
3) The limit for  objects with masses between 0.02 and 0.5~M$_\odot$ becomes less and less 
stringent at higher masses.  
No robust limits can be inferred for objects heavier than 0.2~M$_\odot$ because, with our present sensitivity, 
some models are compatible with a Halo entirely made of such objects. 
Our results do not contradict the positive signal of  the MACHO collaboration (Alcock {\it et al}, 1996b) as is shown in 
figure~\ref{fig_exclupart}.

\section{Conclusions}

The main result of the first phase of the EROS program is that
objects with masses between 10$^{-7}$ and 0.02~M$_\odot$ do not contribute significantly to the Halo dark matter.
The results shown in figure~\ref{fig_exclupart} indicate that there remains a great uncertainty in the proportion of the Halo
 comprised of heavier objects. Further data from the MACHO collaboration and the recently upgraded OGLE and EROS programs should 
clarify the issue in the coming years.

\begin{acknowledgements}
 We are grateful for the support given to our project by the technical staff at ESO La Silla. We thank
T.~Boutreux, R.~Burnage,
 V.~de~Lapparent, F.~Masset, D.~Mouillet, C.~Nitschelm, J.~Poinsignon for their help with the observations.
\end{acknowledgements}

\end{document}